\documentclass[preprint,12pt, a4paper]{elsarticle}

\usepackage{amssymb}

\usepackage{lineno}
\usepackage{subcaption}
\usepackage{hyperref}
\usepackage[dvipsnames]{xcolor}

\def\inst#1{\unskip$^{#1}$}

\journal{SoftwareX}

\begin{document}

\begin{frontmatter}



\title{SPAALUV: Software Package for Automated Analysis of Lung Ultrasound Videos}


\author{Anito Anto\inst{1}}\sep \author{Linda Rose Jimson\inst{1}}\sep \author{Tanya Rose\inst{1}}\sep \author{Mohammed Jafrin\inst{1}}\sep\author{Mahesh Raveendranatha Panicker\inst{2}}

\address{Cochin University of Science and Technology, Cochin, India
\inst{1}}
\address{Indian Institute of Technology Palakkad, India
\inst{2}}

\begin{abstract}

In the recent past with the rapid surge of COVID-19 infections, lung ultrasound has emerged as a fast and powerful diagnostic tool particularly for continuous and periodic monitoring of the lung. There have been many attempts towards severity classification, segmentation and detection of key landmarks in the lung. Leveraging the progress, an automated lung ultrasound video analysis package is presented in this work, which can provide summary of key frames in the video, flagging of the key frames with lung infection and options to automatically detect and segment the lung landmarks. The integrated package is implemented as an open-source web application and available in the link \url{https://github.com/anitoanto/alus-package}. 

\end{abstract}


\end{frontmatter}








\section{Motivation and significance}
Over the last decade, lung ultrasound (LUS) has evolved remarkably as a potent diagnostic tool for a variety of lung disorders \cite{marini2021lung}. With portability, possibility of repeated and dynamic bedside scanning and non-invasive radiation-free nature, LUS is expected to be a defacto scanning methodology in emergency medicine and ambulatory scenarios \cite{jackson2021lung}. In the recent past with COVID-19 outbreak, there have been many developments towards introducing artificial intelligence enabled ultrasound image/video analysis \cite{wang2021deep}-\cite{lee2016lung}, thereby easing the workload of the clinicians involved. In this work, we propose a software package for automated real-time analysis of lung ultrasound videos that can potentially be utilized by even by naive clinicians. The purpose of this package is two fold: 1) to summarize the lung ultrasound videos into a short video of only key frames which helps fast triaging and also enables tele-medicine and 2) to detect and segment the key landmarks in the summarized key frames for easy interpretation.
\begin{figure}[t]
    \centering
    \includegraphics[width=\textwidth]{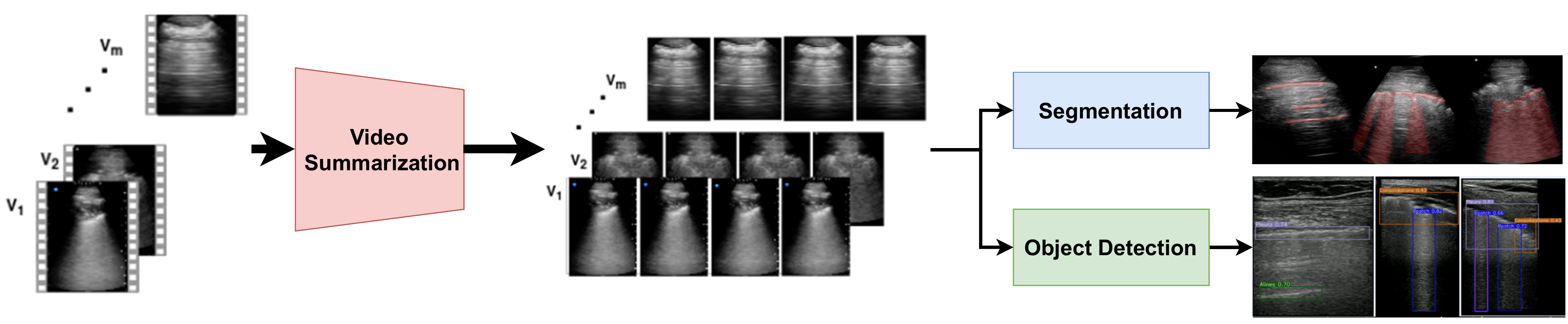}
    \caption{An overview of the proposed package}
    \label{fig1_overview}
\end{figure}

For a large number of ultrasound videos with a massive amount of frames that are difficult to annotate manually, video summarisation relieves the clinicians from the time-consuming burden of selecting the frames that are most relevant to the diagnosis of the patient \cite{zhou2018deep}, \cite{mathews2021unsupervised}. The automated package is delivered as an open-source web application that summarizes the ultrasound videos extracting only the significant frames to generate a compressed version with non-redundant abnormal data. The non-redundant frames can be employed for detection and segmentation of the key landmarks in the summarized key frames, which can be subsequently downloaded by the clinicians for reporting \cite{mojoli2019lung}. A brief overview of the proposed framework is as shown in Fig. \ref{fig1_overview} and more details of the algorithms are available at \cite{mathews2021unsupervised},\cite{joseph2022covecho}.

\section{Software description}
\subsection{Software Architecture}
\begin{figure}[htp]
    \centering
    \includegraphics[width=\textwidth]{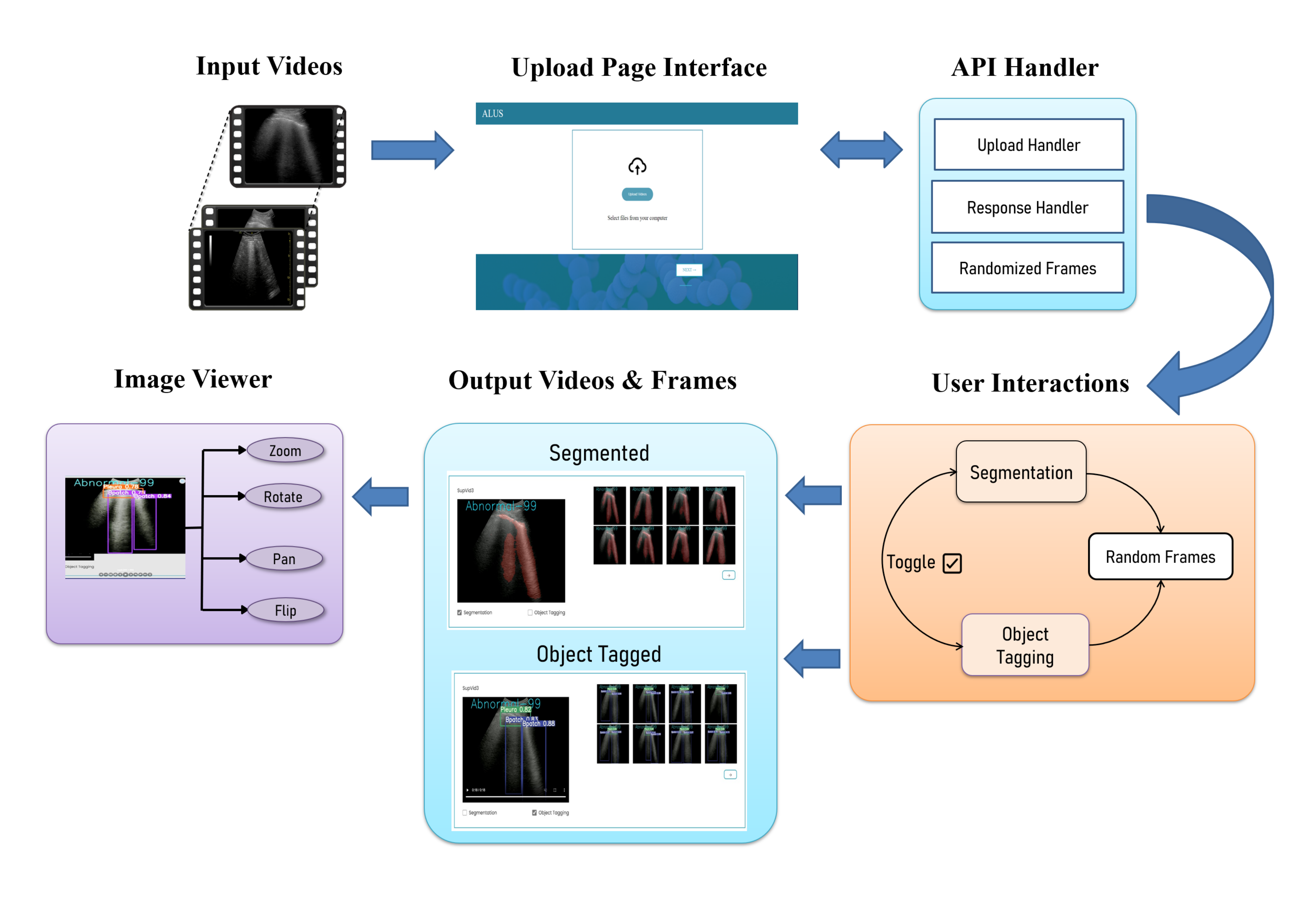}
    \caption{Detailed overview of the front-end flow diagram}
    \label{fig2_front-end}
\end{figure}

The web application consists of a responsive user interface rendered using the React framework in the front-end with a high level descriptive overview of the project.The back end, which encompasses all the non-interface resources, is implemented using the high-performance FastAPI framework.This is integrated with the machine learning algorithms for video summarisation, classification, segmentation, and object tagging \cite{mathews2021unsupervised},\cite{joseph2022covecho}.An overview of the front end and back end are shown in Figures \ref{fig2_front-end} and \ref{fig3_backend} respectively.

The user interface of the web application includes the input controls, navigational components, and informational containers that enable users to easily navigate through the web application and create a positive user experience.The tool comprises of six pages: Home, Upload, Segmentation and Object Tagging, Report Page, About, and Contact Page.Users can navigate between the Home, About, and Contact pages using the navigation bar at the top of the web page.The \textit{upload page} interface contains an upload button that allows the user to upload multiple videos at once. After completing the upload procedure, the user will be acknowledged with the number of videos that have been uploaded to the application. When the user clicks the Next button, the uploaded videos are sent for processing as an input to the model for video summarization.Once the videos are summarised with the key frames, each of the key frames are employed for segmentation and detection of the key landmarks.On the third page, the segmented video is displayed on the left-hand side of the screen, with its corresponding key frames extracted from it on the right.The \textit{segmentation} and \textit{object tagging} checkboxes can be used to toggle between both choices as desired by the user.The segmentation option will be selected by default, and the subsequent web page will show a summarised segmented video with a few random keyframes.The pleura and the abnormalities from the summarised video will be highlighted in a distinct colour from the rest of the image, thus helping in making clinical diagnosis easier.By toggling the segmentation checkbox, we can switch between the normal summarised video and the segmented video with their corresponding frames in order to highlight the visual differences between the original ultrasound video and the summary.The object tagging checkbox will display the object-tagged video with their labelled artefacts enclosed in bounding boxes, accompanied by their confidence scores. The videos can be toggled individually according to the user's preference, and when the choice gets switched, the model does not need to be processed again, allowing the output to be displayed immediately.The random option beneath the frames allows the user to generate random frames each time it is clicked, allowing this feature to be used for each video separately.The tool also allows the user to download the segmented and object-tagged output, allowing them to retrieve valuable information in order to make crucial clinical judgements.

\subsection{Software Functionalities}
\begin{figure}[htp]
    \centering
    \includegraphics[width=\textwidth]{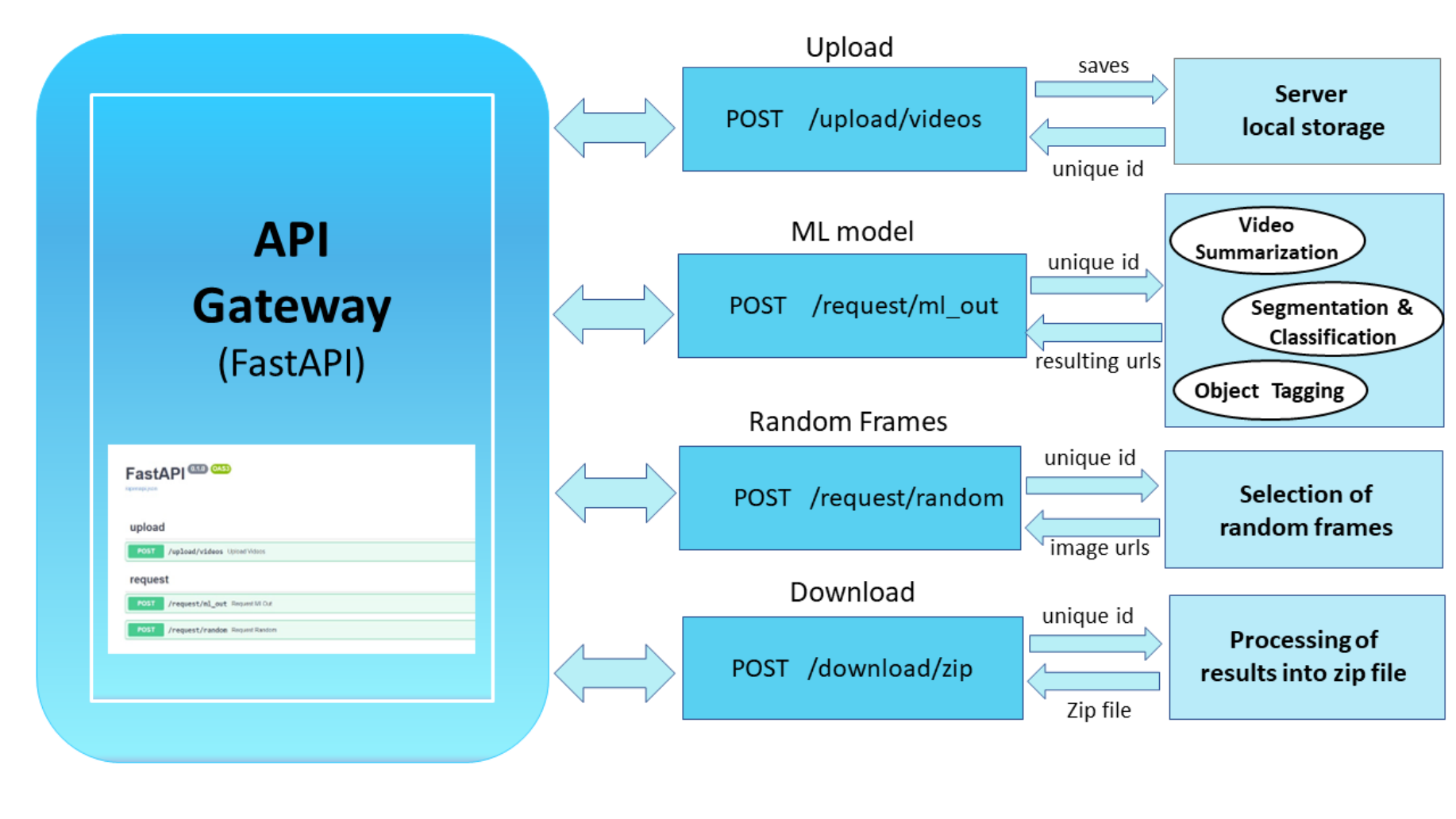}
    \caption{Detailed overview of the back end architecture}
    \label{fig3_backend}
\end{figure}

The lung videos acquired using the obliquely positioned ultrasound probe \cite{gargani2014lung} along the posterior, anterior, and lateral chest walls are uploaded via a POST web request from the front-end interface as shown in Fig. \ref{fig2_front-end}. The Fast API on the Uvicorn server handles the responses according to the type of client-side requests as shown in Fig. \ref{fig3_backend}. In response to the uploaded video, the back end server generates a unique identifier key, which is returned to the client-side. This specific key acts as a conduit for future client-server communication \cite{northwood2018full}. The uploaded videos are processed by triggering a web request by the front end using the unique key identifier to the API endpoint of the server. This ultrasound scan of different areas of a person's lungs serves as an input to the machine learning model.

\newpage
The uploaded ultrasound scan videos with redundant frames may be quite large to process, so the pathologically important key frames are identified using a video summarisation framework. The video summarisation architecture is modelled in the form of ensemble encoders with an LSTM decoder as described in \cite{mathews2021unsupervised}. This unsupervised video summarisation model summarises the uploaded videos by selecting diagnostically important frames crucial for the clinician’s use, thus enhancing its utility in the emergency department as well as in telemedicine. During the pandemic, segmentation of key landmarks and automatic classification of lung video frames into normal and abnormal classes may be useful for a variety of clinical applications and add value in a limited-resource clinical environment. This application also has an underlying AI-driven object detection model with enabled active learning that automatically identifies the important landmarks in the lung. Object tagging is implemented using the \textit{YOLOv5} algorithm, which treats the summarized video as a multi-class problem \cite{joseph2022covecho}. The algorithm displays bounding boxes with their confidence scores for various ultrasound artefacts like A-lines, B-lines, consolidations and landmarks such as pleura, rib and shadow. The segmented and object-tagged videos are saved on the server-side under a unique key. Random frames are extracted from each summarised video. Consequently, the back end sends a response to the client-side request by transmitting a set of URLs for each video and keyframes to be displayed in the front end according to the user preference. This results in the desired summarized video and $8$ random keyframes extracted from each of the videos.

\section{Illustrative Examples}
A step by step guide on the usage of  the software tool depicting the processing of lung ultrasound videos by the automated lung ultrasound package is illustrated below.
\subsection{Upload and Process LUS videos}
\begin{figure}[t]
    \centering
    \includegraphics[width=\textwidth]{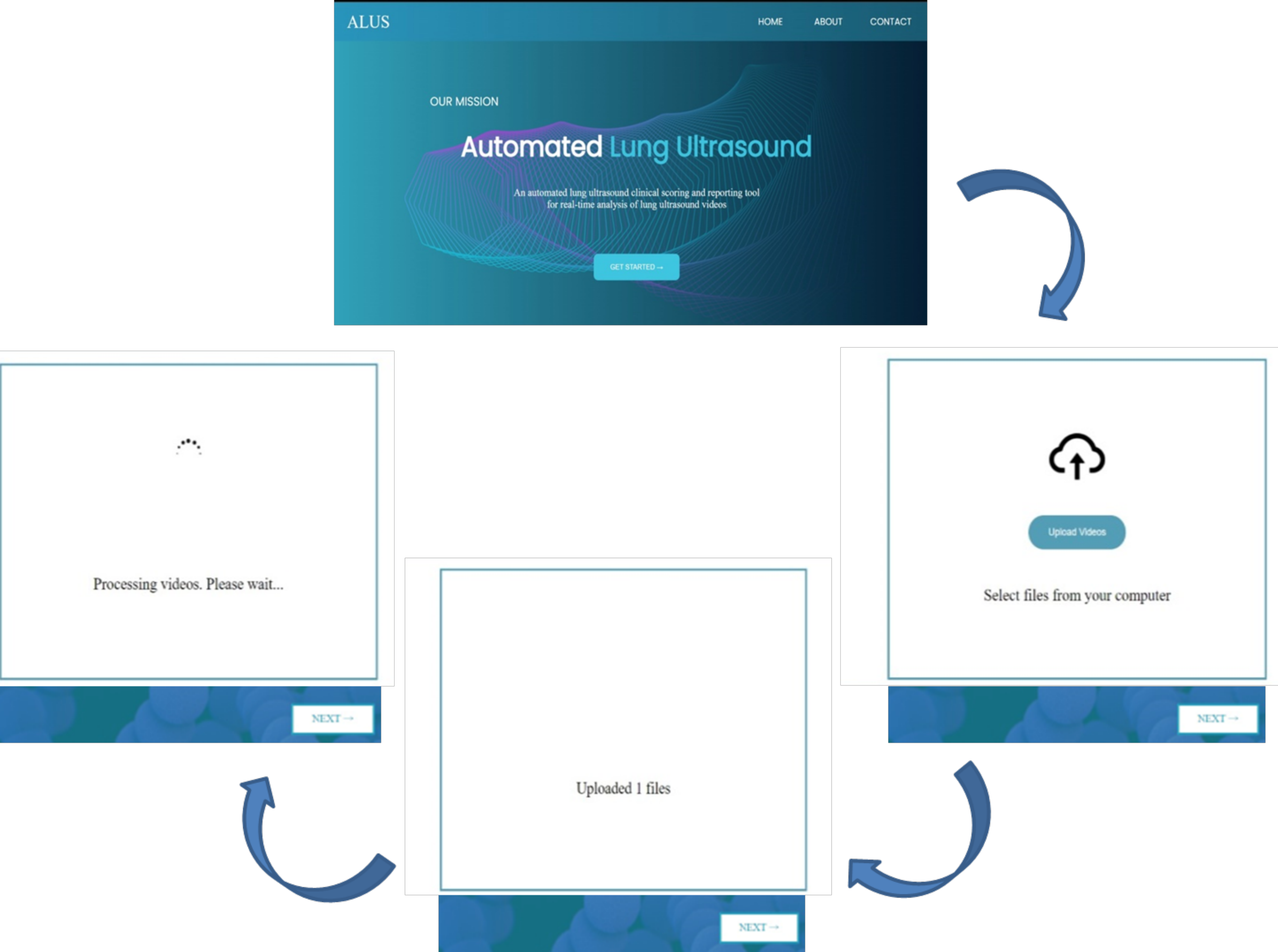}
    \caption{Overview of uploading LUS videos}
    \label{fig4_illustrate(1)}
\end{figure}
The software application begins with a basic overview of the mission of the project. The user must click the \textit{“Get Started”} button to enter the upload page shown in Fig. \ref{fig4_illustrate(1)}. The first step is to upload multiple ultrasound videos obtained using the probe by clicking the \textit{“Upload Videos”} button. Once the user has uploaded videos, it is sent for processing by clicking the \textit{“NEXT”} button. Although we will only upload one video in this example, the procedure is the same for multiple simultaneous analysis of ultrasound scan videos. By refreshing the page, a user can make any necessary changes to the uploaded videos.

\subsection{Generate and Download Results}

\begin{figure}[t]
\vspace{-1cm}
{\begin{subfigure}{\textwidth}
 \includegraphics[width= \textwidth]{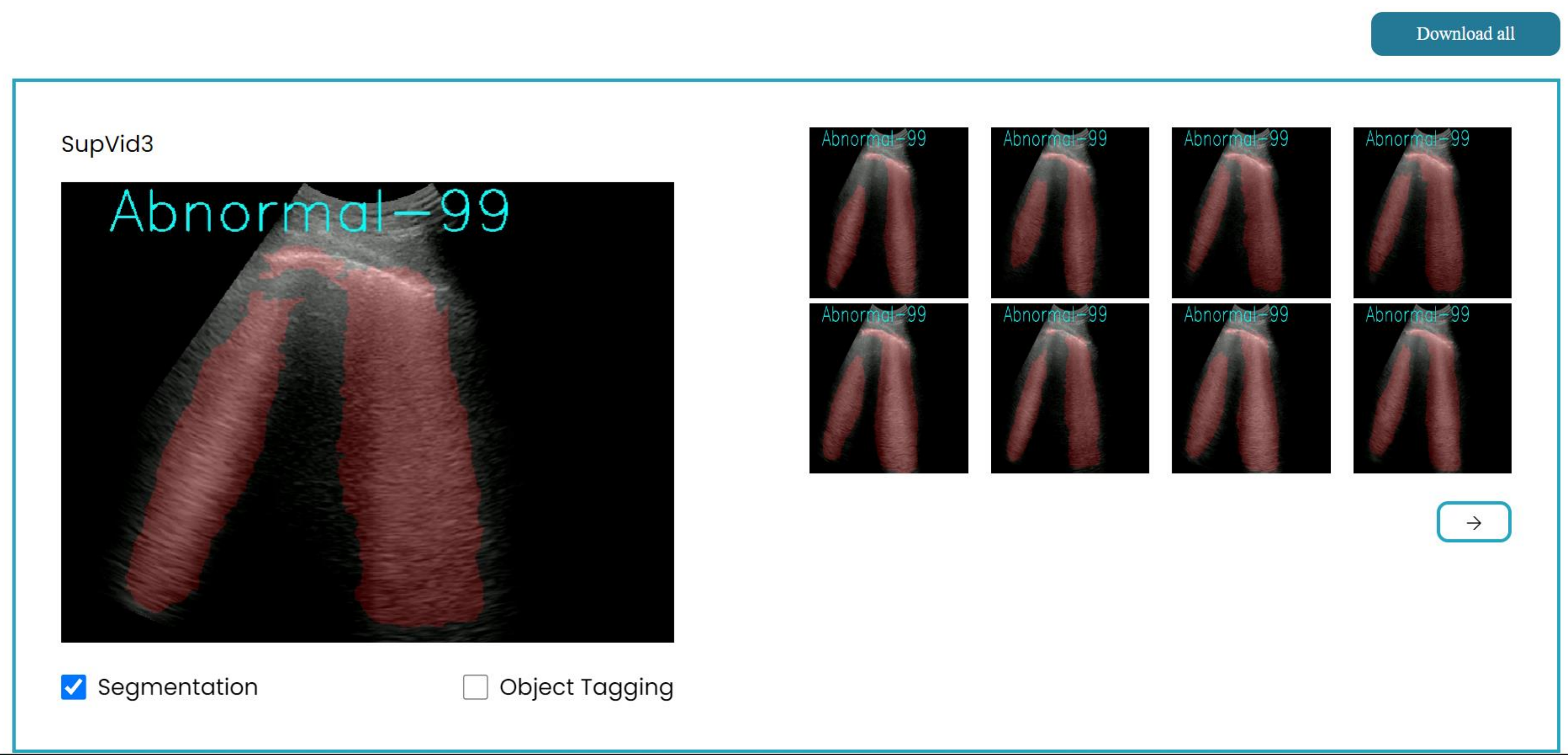}
  \caption{Segmentation Output}
  \label{first}
\end{subfigure}

\begin{subfigure}{\textwidth}
 \includegraphics[width= \textwidth]{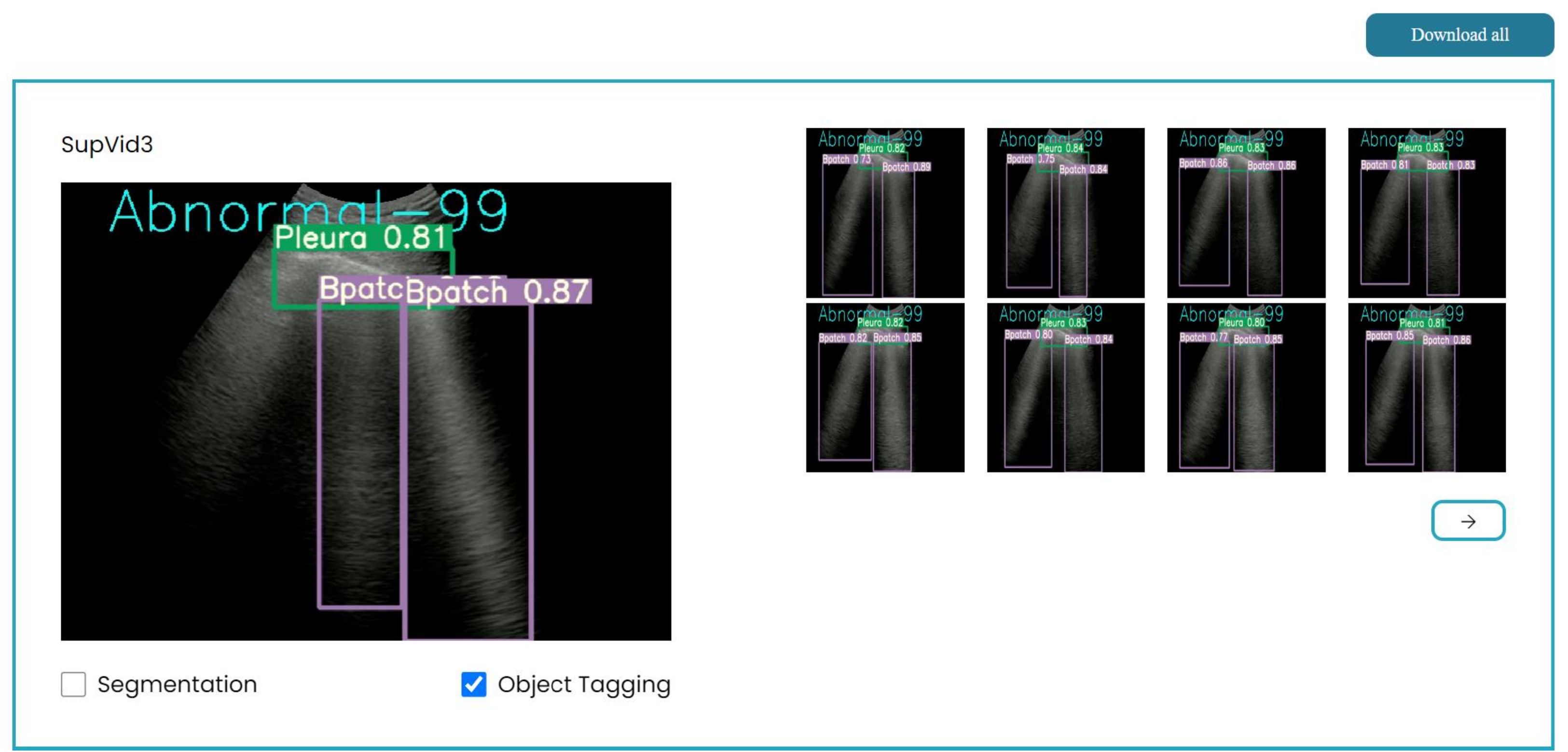}
 \caption{Object-tagging Output}
 \label{second}
\end{subfigure}
\caption{Overview of Result Page}
\label{fig5_illustrate(2)}
}
\end{figure}

The uploaded LUS videos are subjected to segmentation and object tagging algorithms by the automated LUS tool. The segmented and object tagged results are displayed for each ultrasound video separately along with their extracted keyframes as in the Fig.\ref{fig5_illustrate(2)}. Depending on the user's preference, the abnormal areas and the key landmarks in the lung are identified by the \textit{SPAALUV} package. Clinicians are able to conduct a thorough analysis of the report by being able to view each key frame closely in a variety of orientations. For each LUS video, the arrow button on the right side generates an additional set of random keyframes. This software tool also provides the users with a “Download all” button to download all the summarised videos and their processed frames as a zip file for reference purposes in future examinations. The entire demonstration on the usage of the software tool is also provided for user awareness in the \textit{About} page.

\newpage
\section{Impact}
Providing clinicians with a web based easy to use open source software package for automated analysis of lung ultrasound videos is extremely beneficial in the scenarios of pandemic like COVID-19. The package has the capability of summarizing the acquired ultrasound videos in terms of the key frames and also detection and segmentation of the key landmarks in the lung. The tool will enable faster triaging and also become an enabler for telemedicine. Even though, the developed framework employs specific models trained for lung ultrasound diagnostics, the software framework can be easily extended to other ultrasound applications.

\section{Conclusions}
In this work a web based open source software package is developed for automated analysis of lung ultrasound videos. The ultrasound scan videos are summarised using a robust ultrasound video summarisation pipeline, extracting relevant frames and aiding clinicians to interpret the scan data more accurately. The summarised video is further segmented, classified, and object-tagged to help analyse the severity of the lungs efficiently. The proposed automated lung ultrasound tool has been developed and deployed on the cloud for piloting in collaborated hospitals.

\section{Conflict of Interest}
We wish to confirm that there are no known conflicts of interest associated with this publication and there has been no significant financial support for this work that could have influenced its outcome.

\section*{Acknowledgements}
The authors would like to acknowledge the technical contributions from Mr. Roshan P Mathews, Dr. Abhilash R Hareendranathan and Ms. Jinu Joseph in contributing to the development of different machine learning models which are reemployed in the proposed work. The authors are also extremely thankful to the clinical team consisting of Dr. Yale Tung Chen, Dr. Jacob Jaremko, Dr. Kesavadas C and Dr. Kiran Vishnu Narayan for their support in providing with relevant data and clinical support as and when needed. We would like to acknowledge the funding from the Department of Science and Technology - Science and Engineering Research Board (DSTSERB (CVD/2020/000221)) for the CRG COVID-19 funding. The authors are grateful to Compute Canada and the NVIDIA and CDAC for providing the computing resources for the project. 






\begin{thebibliography}{8}

\bibitem{marini2021lung}
Marini, Thomas J., Deborah J. Rubens, Yu T. Zhao, Justin Weis, Timothy P. O’Connor, William H. Novak, and Katherine A. Kaproth-Joslin. "Lung ultrasound: the essentials." Radiology: Cardiothoracic Imaging 3, no. 2 (2021).

\bibitem{jackson2021lung}
Jackson, Karl, Robert Butler, and Avinash Aujayeb. "Lung ultrasound in the COVID-19 pandemic." Postgraduate medical journal 97, no. 1143 (2021): 34-39.

\bibitem{wang2021deep}
Wang, Xi, Hao Chen, Huiling Xiang, Huangjing Lin, Xi Lin, and Pheng-Ann Heng. "Deep virtual adversarial self-training with consistency regularization for semi-supervised medical image classification." Medical image analysis 70 (2021): 102010.

\bibitem{tsai2021automatic}
Tsai, Chung-Han, Jeroen van der Burgt, Damjan Vukovic, Nancy Kaur, Libertario Demi, David Canty, Andrew Wang et al. "Automatic deep learning-based pleural effusion classification in lung ultrasound images for respiratory pathology diagnosis." Physica Medica 83 (2021): 38-45.

\bibitem{mugasa2020adaptive}
Mugasa, Hatwib, Sumeet Dua, Joel EW Koh, Yuki Hagiwara, Oh Shu Lih, Chakri Madla, Pailin Kongmebhol, Kwan Hoong Ng, and U. Rajendra Acharya. "An adaptive feature extraction model for classification of thyroid lesions in ultrasound images." Pattern Recognition Letters 131 (2020): 463-473.

\bibitem{lee2016lung}
Lee, Francis Chun Yue. "Lung ultrasound—a primary survey of the acutely dyspneic patient." Journal of intensive care 4, no. 1 (2016): 1-13.

\bibitem{ouahabi2021deep}
Ouahabi, Abdeldjalil, and Abdelmalik Taleb-Ahmed. "Deep learning for real-time semantic segmentation: Application in ultrasound imaging." Pattern Recognition Letters 144 (2021): 27-34.

\bibitem{zhou2021multi}
Zhou, Yue, Houjin Chen, Yanfeng Li, Qin Liu, Xuanang Xu, Shu Wang, Pew-Thian Yap, and Dinggang Shen. "Multi-task learning for segmentation and classification of tumors in 3D automated breast ultrasound images." Medical Image Analysis 70 (2021): 101918.

\bibitem{wang2021deep1}
Wang, Yu, Xinke Ge, He Ma, Shouliang Qi, Guanjing Zhang, and Yudong Yao. "Deep learning in medical ultrasound image analysis: a review." IEEE Access 9 (2021): 54310-54324.

\bibitem{litjens2017survey}
Litjens, Geert, Thijs Kooi, Babak Ehteshami Bejnordi, Arnaud Arindra Adiyoso Setio, Francesco Ciompi, Mohsen Ghafoorian, Jeroen Awm Van Der Laak, Bram Van Ginneken, and Clara I. Sánchez. "A survey on deep learning in medical image analysis." Medical image analysis 42 (2017): 60-88.

\bibitem{zhou2018deep}
Zhou, Kaiyang, Yu Qiao, and Tao Xiang. "Deep reinforcement learning for unsupervised video summarization with diversity-representativeness reward." In Proceedings of the AAAI Conference on Artificial Intelligence, vol. 32, no. 1. 2018.

\bibitem{mathews2021unsupervised}
Mathews, Roshan P., Mahesh Raveendranatha Panicker, Abhilash R. Hareendranathan, Yale Tung Chen, Jacob L. Jaremko, Brian Buchanan, Kiran Vishnu Narayan, and Greeta Mathews. "Unsupervised multi-latent space reinforcement learning framework for video summarization in ultrasound imaging." arXiv preprint arXiv:2109.01309 (2021).

\bibitem{mojoli2019lung}
Mojoli, Francesco, Bélaid Bouhemad, Silvia Mongodi, and Daniel Lichtenstein. "Lung ultrasound for critically ill patients." American journal of respiratory and critical care medicine 199, no. 6 (2019): 701-714.

\bibitem{gargani2014lung}
Gargani, Luna, and Giovanni Volpicelli. "How I do it: lung ultrasound." Cardiovascular ultrasound 12, no. 1 (2014): 1-10.

\bibitem{northwood2018full}
Northwood, Chris. The Full Stack Developer: Your Essential Guide to the Everyday Skills Expected of a Modern Full Stack Web Developer. Apress, 2018.

\bibitem{joseph2022covecho}
Joseph, Jinu, Mahesh Raveendranatha Panicker, Yale Tung Chen, Kesavadas Chandrasekharan, Vimal Chacko Mondy, Anoop Ayyappan, Jineesh Valakkada, and Kiran Vishnu Narayan. "covEcho Resource constrained lung ultrasound image analysis tool for faster triaging and active learning." arXiv preprint arXiv:2206.10183 (2022).

\end{thebibliography}
\newpage
\section*{Current executable software version}
\label{currSoftware}
\begin{table}[!h]
\begin{tabular}{p{5.5cm}|p{7.5cm}}
\hline
Current software version & v1.0 \\
\hline
Permanent link to code/repository used for this version  &  $https://github.com/anitoanto/alus-package$ \\
\hline
Legal Software License & MIT License \\
\hline
Software code languages, tools, and services used & Python 3.9, FastAPI, React, Node LTS \\
\hline
Installation requirements \& dependencies & backend:python 3.9.x, packages in requirements.txt file

frontend:node 16.16.0 LTS, npm, packages in package.json file, browser

specification:
good hardware for ML
processing, preferably intel i5 8gen or above
\\
\hline-
If available Link to developer documentation/manual &
$https://github.com/anitoanto/alus-package/blob/main/README.md$ \\
\hline
Support email for questions & $anitoanto07@gmail.com$ \\
\hline
\end{tabular}
\end{table}

\end{document}